\documentclass[twocolumn]{aastex63}
\usepackage[normalem]{ulem}
\usepackage{graphicx}
\usepackage{dcolumn}
\usepackage{bm}
\usepackage{color}
\usepackage{array}
\usepackage{multirow}
\usepackage{amsmath}
\usepackage{subfigure}

\newcommand{\f}{\frac}

\newcommand{\n}{\nonumber}
\newcommand{\p}{\partial}

\renewcommand\arraystretch{1.3}

\begin{document}
\title{Cosmological Parameter Estimation for Dynamical Dark Energy Models with Future Fast Radio Burst Observations}

\author{Ze-Wei Zhao}
\affiliation{Department of Physics, College of Sciences, $\&$ Ministry of Education's Key Laboratory of Data Analytics and Optimization for Smart Industry, Northeastern University, Shenyang 110819, China; zhangxin@mail.neu.edu.cn}

\author{Zheng-Xiang Li}
\affiliation{Department of Astronomy, Beijing Normal University, Beijing 100875, China; zxli918@bnu.edu.cn, gaohe@bnu.edu.cn}

\author{Jing-Zhao Qi}
\affiliation{Department of Physics, College of Sciences, $\&$ Ministry of Education's Key Laboratory of Data Analytics and Optimization for Smart Industry, Northeastern University, Shenyang 110819, China; zhangxin@mail.neu.edu.cn}

\author{He Gao}
\affiliation{Department of Astronomy, Beijing Normal University, Beijing 100875, China; zxli918@bnu.edu.cn, gaohe@bnu.edu.cn}

\author{Jing-Fei Zhang}
\affiliation{Department of Physics, College of Sciences, $\&$ Ministry of Education's Key Laboratory of Data Analytics and Optimization for Smart Industry, Northeastern University, Shenyang 110819, China; zhangxin@mail.neu.edu.cn}

\author{Xin Zhang}
\affiliation{Department of Physics, College of Sciences, $\&$ Ministry of Education's Key Laboratory of Data Analytics and Optimization for Smart Industry, Northeastern University, Shenyang 110819, China; zhangxin@mail.neu.edu.cn}

\begin{abstract}
Fast radio bursts (FRBs) are a mysterious astrophysical phenomenon of bright pulses emitted at radio frequencies, and they are expected to be frequently detected in the future. The dispersion measures of FRBs are related to cosmological parameters, thus FRBs have the potential to be developed into a new cosmological probe if their data can be largely accumulated in the future. In this work, we study the capability of future FRB data to improve cosmological parameter estimation in two dynamical dark energy models. We find that the simulated FRB data can break the parameter degeneracies inherent in the current cosmic microwave background (CMB) data. Therefore, the combination of the CMB and FRB data can significantly improve the constraints on the Hubble constant and dark energy parameters, compared to those using CMB or FRB alone. If 10,000 FRB events with known redshifts are detected in the future, they would perform better than the baryon acoustic oscillation (BAO) data in breaking the parameter degeneracies inherent in the CMB data. We also find that the combination of FRB and gravitational-wave (GW) standard siren data provides an independent low-redshift probe to verify the results from the CMB and BAO data. For the data combination of CMB, GW, and FRB, it is found that the main contribution to the constraints comes from the CMB and GW data, but the inclusion of the FRB data still can evidently improve the constraint on the baryon density.
\end{abstract}



\section{Introduction} \label{sec:intro}

The late-time cosmic accelerated expansion discovered by \citet{Riess:1998cb} and \citet{Perlmutter:1998np} cannot be realized in a universe governed by general relativity with only barotropic and pressureless fluids. To realize the acceleration in the late universe, one needs to modify general relativity at the cosmological scale or introduce a new component with negative pressure, called dark energy (DE). The cosmic microwave background (CMB) anisotropies data measured by the Planck satellite \citep{Aghanim:2018eyx} favor the $\Lambda$ cold dark matter ($\Lambda$CDM) model with the DE provided by a cosmological constant ($\Lambda$), which is usually regarded as the standard model of cosmology \citep{Bahcall}. However, the $\Lambda$CDM model suffers from the cosmological constant problem \citep{Weinberg}, so the proposal of dynamical dark energy has also been widely studied \citep{Joyce}.

The CMB data alone can only constrain the cosmological parameters at high precision for the base $\Lambda$CDM model, but they cannot provide precise estimations for the extra parameters if the model is extended to include new physics; in particular, there usually exist strong degeneracies between these parameters \citep{Akrami}. Since the CMB observation is the measurement of the early universe, low-redshift observations like the baryon acoustic oscillation (BAO) observation are usually employed as complements to break the parameter degeneracies \citep{Beutler:2011hx,Ross:2014qpa,Alam:2016hwk}. It should be pointed out that although the current BAO measurements come from the galaxy redshift surveys for late universe, the BAO and CMB observations actually share the same standard ruler of the comoving scale of sound horizon formed in the early universe. Thus, developing other independent and precise low-redshift cosmological probes to verify the results from CMB+BAO is of great interest and importance.

Recently, a class of bright pulses with millisecond-duration at radio frequencies, named fast radio bursts (FRBs), has been detected \citep{2007Sci...318..777L,2013Sci...341...53T,2015MNRAS.447..246P,2016PASA...33...45P}. Although the FRBs' specific progenitors are still unknown, their locations are considered to be extragalactic, for the dispersion measures (DMs) of FRBs greatly exceeding the maximum Galactic expectations. Indeed, the cosmological redshifts and the host galaxies of several FRBs have been recently identified \citep{2016ApJ...833..177S,2016Natur.531..202S,2017Natur.541...58C,2017ApJ...834L...8M,2017ApJ...834L...7T,2019arXiv190611476B,2019arXiv190701542R}, including both repeating and nonrepeating events. It has long been proposed that a sufficiently large sample of FRBs with redshift detection could be used to place constraints on cosmological parameters through the DM--redshift relation \citep{2014ApJ...788..189G,2014PhRvD..89j7303Z}. Current FRB observations suggest a sufficiently high all-sky FRB rate of $\sim10^{3}-10^{4}$ per day \citep{2019ARA&A..57..417C,2019A&ARv..27....4P}. Current and upcoming surveys, such as the Canadian Hydrogen Intensity Mapping Experiment (CHIME) telescope and its FRB search backend \citep{2018ApJ...863...48C}, and especially the Square Kilometre Array (SKA) project \citep{2015aska.confE..55M}, are expected to detect about 1000 or more FRB events every day \citep{Fialkov:2017qoz}, making the constraint of cosmological parameters feasible.

In the literature, the cosmological parameter estimation from FRB was pioneered by using the possible association of FRBs and gamma-ray bursts to measure the intergalactic medium portion of the baryon mass fraction of the universe \citep{2014ApJ...783L..35D} and to conduct cosmography \citep{2014ApJ...788..189G}, and actually FRBs are expected to be developed into a promising tool to study the expansion history of the universe; see a series of works, e.g. breaking the cosmological parameter degeneracies by combining the FRB data with the BAO data \citep{2014PhRvD..89j7303Z} and the type Ia supernova (SN) data \citep{Jaroszynski:2018vgh}, directly constraining cosmological parameters by introducing a slope parameter \citep{2016ApJ...830L..31Y}, extracting cosmological distance information by considering the effects of systematic uncertainties \citep{Kumar:2019qhc}, acting as a cosmological probe from strongly lensed repeating FRBs \citep{2018NatCo...9.3833L,Liu:2019jka}, measuring the cosmic proper distance \citep{2017A&A...606A...3Y},
and improving the constraints on baryon density compared to the current data set \citep{2018ApJ...856...65W}. However, the combination of FRB and the current most precise cosmological probe, the CMB observation, has still not been deeply studied.

Additionally, an interesting and heuristic idea is the gravitational-wave (GW)/FRB association proposed by \citet{2018ApJ...860L...7W}, in which the authors noticed that the combination of luminosity distance $d_{\rm L}$ from GW and DM of FRB would be very helpful for cosmological tests. Inspired by this idea, \citet{Cai:2019cfw} studied cosmic anisotropy with GW/FRB association and \citet{2019ApJ...876..146L} introduced a cosmology-independent estimate of the fraction of baryon mass in the intergalactic medium (IGM). It is suggested that the correlations of cosmological parameters in DM and in $d_{\rm L}$ are rather different, even opposite, because DM is proportional to the Hubble constant $H_0$ whereas $d_{\rm L}$ is inversely proportional to $H_0$. This fact may also be very helpful in cosmological parameter estimation and need to be further studied.

The GWs detected by the Laser Interferometer Gravitational-Wave Observatory (LIGO) are astronomical low-redshift events produced by the mergers of binary black hole or binary neutron-star (BNS) systems, which are fully independent of the high-redshift CMB observation. The advantage of GWs is that the absolute distance information of the source can be directly extracted from the GW signal, which discards the distance ladder method to calibrate between different astronomical processes. The GW events observed by next-generation ground-based GW detectors, such as the Einstein Telescope (ET), combined with independent electromagnetic observations, can lead to a true distance--redshift relation, which can be used to study cosmology. Thus, such GW sources are often dubbed ``standard sirens" \citep{Schutz1986,Holz:2005df}, and are expected to become a new precise cosmological probe \citep{Sathyaprakash:2009xs,Sathyaprakash:2009xt,Zhao:2010sz,Li:2013lza,Cai:2016sby,Cai:2017aea,Chen:2017rfc,Wang:2018lun,Wang:2019tto,Feeney:2018mkj,Li:2019ajo,Zhang:2019ylr,Zhang:2019loq,Zhang:2018byx,Zhang:2019ple,Jin:2020hmc,Zhao:2019gyk}. In particular, it is found that the combination of GW and other cosmological probes, such as CMB and BAO, is greatly helpful for breaking the parameter degeneracies \citep{Wang:2018lun,Wang:2019tto,Zhang:2019ylr,Zhang:2019loq,Zhang:2018byx,Zhang:2019ple,Jin:2020hmc,Zhao:2019gyk}.

In this paper, we wish to study the capability of future FRB data to break the cosmological parameter degeneracies inherent in the CMB data from Planck and the GW data from ET.

This paper is organized as follows. In Section \ref{sec:Method}, we briefly introduce the methods for simulating the FRB data and the standard siren data. We show the constraint results and provide relevant discussions in Section \ref{sec:Result}. We present our conclusions in Section \ref{sec:con}.


\section{Methods and data}\label{sec:Method}

\subsection{Simulation of FRBs}\label{21}
In this work, we study two dynamical dark energy models: the $w$CDM model with the equation of state (EoS) of DE being a constant, $w(z)=p_{\rm de}(z)/\rho_{\rm de}(z) = w$, and the Chevallier--Polarski--Linder (CPL) model with the EoS of DE parameterized by the form, $w(z)=w_{\rm{0}}+w_{\rm{a}}z/(1+z)$ \citep{Chevallier:2000qy,Linder:2002et}. According to the Friedmann equation, the dimensionless Hubble parameter in a flat universe is given by
\begin{align}\label{eq:dL0}
E^2(z)=\frac{H^2(z)}{H_0^2}=\,&(1 - {\Omega _{\rm m}})\exp \left[3\int_0^z {\frac{{1 + w(z')}}{{1 + z'}}} d z'\right] \nonumber\\
&+{\Omega _{\rm m}}{(1 + z)^3},
\end{align}
where $H(z)$ is the Hubble parameter, $H_0=100h\, {\rm km\, s^{-1}\, Mpc^{-1}}$ is the Hubble constant, and ${\Omega _{\rm m}}$ is the present-day matter density parameter.


In order to generate a mock sample of future detectable FRBs, we first need to assume a redshift distribution of FRBs. Until now, the progenitors of FRBs have not generally been identified, so the real redshift distribution of FRBs is still unknown. Therefore, following \citet{2019ApJ...876..146L}, we phenomenologically assume that the sources of FRBs {have a constant comoving number density},
\begin{equation}\label{eq4}
N_{\rm{const}}(z)=\mathcal{N}_{\rm{const}}\frac{{d^2_{\rm C}}(z)}{H(z)(1+z)}e^{-{d^2_{\rm{L}}}(z)/[2{d^2_{\rm{L}}}(z_{\rm cut})]},
\end{equation}
where $\mathcal{N}_{\rm{const}}$ is a normalization factor and $d_{\rm C}$ is the comoving distance at redshift $z$. We also include a Gaussian cutoff at redshift $z_{\rm cut}=1$ to represent the decrease of the detected FRBs beyond it due to the instrumental signal-to-noise threshold effect.

{The observed DM of an FRB is quantified by the arrival time delay between the highest and lowest frequencies of the pulse, which consists of contributions from the FRB's host galaxy, IGM, and the Milky Way} \citep{2013Sci...341...53T,2014ApJ...783L..35D}, i.e.,
\begin{equation}\label{eq2}
\rm{DM}_{\rm{obs}}=\rm{DM}_{\rm{host}}+\rm{DM}_{\rm{IGM}}+\rm{DM}_{\rm{MW}}.
\end{equation}
Among these components, $\rm{DM}_{\rm{IGM}}$ is related to cosmology, and its average value can be expressed as
\begin{equation}\label{eq3}
\langle\mathrm{DM}_{\mathrm{IGM}}\rangle=\frac{3cH_0\Omega_bf_{\mathrm{IGM}}}{8\pi G m_{\mathrm{p}}}\int_0^z\frac{\chi(z')(1+z')dz'}{E(z')},
\end{equation}
where
\begin{equation}
\chi(z)=Y_{\rm H}\chi_\mathrm{{e,H}}(z)+\frac{1}{2}Y_{\rm He}\chi_\mathrm{{e,He}}(z).
\end{equation}
In this expression, $\Omega_\mathrm{b}$ is the present-day baryon density parameter, $f_{\mathrm{IGM}}\simeq 0.83$ is the fraction of baryon mass in the IGM \citep{2012ApJ...759...23S}, $m_{\mathrm{p}}$ is the mass of proton, $Y_{\rm H}=3/4$ and $Y_{\rm He}=1/4$ are the mass fractions of hydrogen and helium, respectively, and $\chi_\mathrm{{e,H}}$ and $\chi_\mathrm{{e,He}}$ are the ionization fractions for hydrogen and helium, respectively. Since both hydrogen and helium are fully ionized at $z<3$, we take $\chi_\mathrm{{e,H}}=\chi_\mathrm{{e,He}}=1$ \citep{2006ARA&A..44..415F}.

From Eq. (\ref{eq2}), $\rm{DM}_{\rm{IGM}}$ can be measured for an FRB if $\rm{DM}_{\rm{obs}}$, $\rm{DM}_{\rm{host}}$, and $\rm{DM}_{\rm{MW}}$ could be determined. Thus, the total uncertainty of $\rm{DM}_{\rm{IGM}}$ is
\begin{equation}\label{eq6}
\sigma_{\rm{DM}_{\rm{IGM}}}=\left[\sigma_{\rm obs}^{2}+\sigma_{\rm MW}^{2}+\sigma_{\rm IGM}^{2}
+\left(\frac{\sigma_{\rm host}}{1+z}\right)^{2} \right]^{1/2}.
\end{equation}
The observational uncertainty $\sigma_{\rm {obs}}=1.5~{\rm {pc~cm^{-3}}}$ is adopted from the average value of the released data \citep{2016PASA...33...45P}. According to the Australia Telescope National Facility pulsar catalog \citep{Manchester:2004bp}\footnote{http://www.atnf.csiro.au/research/pulsar/psrcat/}, the average uncertainty of $\rm{DM}_{\rm{MW}}$ for the sources at high Galactic latitude is about $10~{\rm {pc~cm^{-3}}}$. The uncertainty $\sigma_{\rm {IGM}}$ describes the deviation of an individual event from the mean $\rm{DM}_{\rm{IGM}}$, due to the inhomogeneity of the baryon matter in the IGM. {Here we use the following step formula \citep{2019ApJ...876..146L},
\begin{equation}\label{sigmaigm}
\sigma_{\rm IGM}=\begin{cases}\frac{52-45z-263z^2+21z^3+582z^4}{1-4z+7z^2-7z^3+5z^4}, &z\leq 1.03,\\
 {-416+270z+480z^2}\\{~~~~~~~~~~+23z^3-162z^4}, &1.03<z\leq 1.3,\\
38\arctan [0.6z+1]+17, &z>1.3,
\end{cases}
\end{equation}
which is fitted from the simulations \citep{FaucherGiguere:2011cy,McQuinn:2013tmc}, to characterize the uncertainty of $\rm{DM}_{\rm{IGM}}$.} It is difficult to estimate $\sigma_{\rm{host}}$, because it generally depends on the individual properties of an FRB, such as the type of the host galaxy, the location of FRB in the host galaxy, and the near-source plasma. We take $\sigma_{\rm{host}} = 30 ~{\rm {pc~cm^{-3}}}$ as the uncertainty of the ${\rm DM_{host}}$.

According to the FRB event rate estimated from the current detections, the future midfrequency component of SKA is likely to detect $\sim 10^3 \, {\rm sky}^{-1}{\rm day}^{-1}$ of FRBs \citep{Fialkov:2017qoz}. Further assuming that 5\% of the detected FRBs can be sufficiently localized to confirm their host galaxies and considering the bright emission lines of the host galaxy for the repeating FRB 121102 \citep{2017ApJ...834L...7T}, we assume $\sim 10$ redshifts of FRB host galaxies per night can be detected by optical telescopes \citep{2018ApJ...856...65W}. Thus, we consider a normal expected scenario with the event number of FRBs $N_{\rm FRB}=1000$ and an optimistic scenario with $N_{\rm FRB}=10,000$ for a few years.

\subsection{Simulation of Standard Sirens}
To simulate the standard siren data from ET, we need to also assume the redshift distribution of GWs~\citep{Zhao:2010sz,Cai:2016sby},
\begin{equation}
P(z)\propto \frac{4\pi {d^2_{\rm C}}(z)R(z)}{H(z)(1+z)},
\label{equa:pz}
\end{equation}
where $R(z)$ is the time evolution of the burst rate with the form \citep{Schneider:2000sg,Cutler:2009qv,Cai:2016sby}
\begin{eqnarray}
R(z)=\left\{
\begin{array}{rcl}
1+2z, & z\leq 1, \\
\frac{3}{4}(5-z), & 1<z<5, \\
0, & z\geq 5.
\end{array} \right.
\label{equa:rz}
\end{eqnarray}

The GW signal $h(t)$ in general relativity consists of two polarizations and can be formulated with the antenna pattern functions $F$ as
\begin{equation}
h(t)=F_+(\theta, \phi, \psi)h_+(t)+F_\times(\theta, \phi, \psi)h_\times(t),
\end{equation}
where $\psi$ is the polarization angle and ($\theta$, $\phi$) are the location angles of the source in the detector frame. The antenna pattern functions of one Michelson-type interferometer of ET are \citep{Zhao:2010sz}
\begin{align}
F_+^{(1)}(\theta, \phi, \psi)=&~~\frac{{\sqrt 3 }}{2}\Big[\frac{1}{2}(1 + {\cos ^2}\theta )\cos (2\phi )\cos (2\psi ) \nonumber\\
                              &~~- \cos \theta \sin (2\phi )\sin (2\psi )\Big],\nonumber\\
F_\times^{(1)}(\theta, \phi, \psi)=&~~\frac{{\sqrt 3 }}{2}\Big[\frac{1}{2}(1 + {\cos ^2}\theta )\cos (2\phi )\sin (2\psi ) \nonumber\\
                              &~~+ \cos \theta \sin (2\phi )\cos (2\psi )\Big].
\label{equa:F}
\end{align}
Three interferometers of ET have an azimuthal difference of $60^\circ$ with each other, so the other two antenna pattern functions are $F_{+,\times}^{(2)}(\theta, \phi, \psi)=F_{+,\times}^{(1)}(\theta, \phi+2\pi/3, \psi)$ and $F_{+,\times}^{(3)}(\theta, \phi, \psi)=F_{+,\times}^{(1)}(\theta, \phi+4\pi/3, \psi)$.

It is convenient to analyze GW data in the Fourier space. By using the stationary phase approximation, we can obtain the Fourier transform for a GW signal,
\begin{eqnarray}
\mathcal{H}(f)=\mathcal{A}f^{-7/6}e^{i\Psi},
\label{equa:hf}
\end{eqnarray}
where $\mathcal{A}$ is the amplitude in the Fourier space,
\begin{align}
\mathcal{A}=&~~\frac{1}{d_{\rm L}}\sqrt{F_+^2(1+\cos^2\iota)^2+4F_\times^2\cos^2\iota}\nonumber\\
            &~~\times \sqrt{5\pi/96}\pi^{-7/6}\mathcal{M}_{\rm c}^{5/6}.
\label{equa:A}
\end{align}
Here, the luminosity distance is given by
\begin{eqnarray}\label{eq:dL0}
d_{\rm L}(z)=\frac{1+z}{H_0}\int_{0}^{z}\frac{c\,dz^{\prime}}{E(z')},
\end{eqnarray}
$\mathcal{M}_{\rm c}=(1+z)M \eta^{3/5}$ is the observed \emph{chirp mass}, $M=m_1+ m_2$ is the total mass of coalescing binary, $m_1$ and $m_2$ are the masses of black holes (BH) or neutron stars (NS), and $\eta=m_1 m_2/M^2$ is the symmetric mass ratio. The definition of the function $\Psi$ can refer to \citet{Sathyaprakash:2009xs} and \citet{Zhao:2010sz}. The parameter $\iota$ denotes the inclination angle between the direction of binary's orbital angular momentum and the line of sight. The corresponding redshifts of GWs are measured by identifying their electromagnetic counterparts, such as the short gamma ray bursts (SGRBs). Experimentally, SGRBs are supposed to be strongly beamed, which implies that the observation requires them to be oriented
nearly face-on (i.e., $\iota\simeq 0$). Computationally, when we apply the Fisher matrix to the GW waveform, averaging over $\iota$ and $\psi$ with the maximal inclination $\iota=20^\circ$ is roughly equal to taking $\iota=0$ \citep{Li:2013lza}. Thus, for the fiducial values of the simulated GW sources, we take $\iota=0$ and the dependence for $\psi$ drops out of the expression for the antenna pattern function.

Whether a signal is confirmed as a GW detection is determined by the signal-to-noise ratio (SNR) measured by the detector. The combined SNR for the network of ET is
\begin{equation}
\rho=\sqrt{\sum\limits_{i=1}^{3}(\rho^{(i)})^2},
\label{euqa:rho}
\end{equation}
where $\rho^{(i)}=\sqrt{\left\langle \mathcal{H}^{(i)},\mathcal{H}^{(i)}\right\rangle}$ is the SNR of the $i$th interferometer, with the inner product being defined as
\begin{equation}
\left\langle{a,b}\right\rangle=4\int_{f_{\rm lower}}^{f_{\rm upper}}\frac{\tilde a(f)\tilde b^\ast(f)+\tilde a^\ast(f)\tilde b(f)}{2}\frac{df}{S_{\rm n}(f)},
\label{euqa:product}
\end{equation}
where ``$\sim$" denotes the Fourier transform of the function and $S_{\rm n}(f)$ is the one-sided noise power spectral density. For simplicity, we limit the integral interval within $[1\,{\rm Hz}, 2f_{\rm LSO}]$ with $f_{\rm LSO}=1/[6^{3/2} 2\pi (1+z)M]$, and take  the fitting formula $S_{\rm n}(f)$ of ET from~\citet{Zhao:2010sz}.



As a preliminary forecast, we use the Fisher information matrix method to estimate the instrumental error on the measurement of $d_{L}$ as
\begin{eqnarray}
\sigma_{d_{\rm L}}^{\rm inst}\simeq \sqrt{\left\langle\frac{\partial \mathcal H}{\partial d_{\rm L}},\frac{\partial \mathcal H}{\partial d_{\rm L}}\right\rangle^{-1}}.
\end{eqnarray}
With the GW waveform in Eq. (\ref{equa:hf}) and assuming that $d_{\rm L}$ is independent of other parameters, we have $\sigma_{d_{\rm L}}^{\rm inst}\propto d_{\rm L}/ \rho$. As described above, the fiducial value of the inclination angle is set to be 0. However, there is a strong degeneracy between $\iota$ and $d_{\rm L}$ in the real analysis, so the impact of $\iota$ should be taken into account when we estimate the practical instrumental error of $d_{\rm L}$. The maximal effect of $\iota$ on the SNR is a factor of 2 (between the source being face-on, $\iota=0$, and edge-on, $\iota=\pi/2$), thus we indeed consider the instrumental error accounting for the degeneracy between $\iota$ and $d_{\rm L}$ \citep{Li:2013lza},
\begin{equation}
\sigma_{d_{\rm L}}^{\rm inst}\simeq \frac{2d_{\rm L}}{\rho}.
\label{sigmainst}
\end{equation}
There is also weak-lensing error caused by the gravity effect of galaxies, which can be approximated as $\sigma_{d_{\rm L}}^{\rm lens}$ = $0.05z d_{\rm L}$. Thus, the total error of $d_{L}$ is
\begin{align}
\sigma_{d_{\rm L}}=&\sqrt{(\sigma_{d_{\rm L}}^{\rm inst})^2+(\sigma_{d_{\rm L}}^{\rm lens})^2} \nonumber\\
            =&\sqrt{\left(\frac{2d_{\rm L}}{\rho}\right)^2+(0.05z d_{\rm L})^2}.
\label{sigmadl}
\end{align}

Following the estimate in \citet{Sathyaprakash:2009xt} and \citet{Cai:2016sby}, we simulate 1000 standard siren events detected by ET during a 10 yr run and take the ratio of BH--NS (i.e. the binary system of a black hole and a neutron star) and BNS events to be 0.03. We also set the mass distributions in the interval [1,2] $M_{\odot}$ for NS and [3,10] $M_{\odot}$ for BH, where $M_{\odot}$ denotes the solar mass.

Although the instrumental error of $d_{L}$ is estimated by applying the Fisher matrix method, we use the Markov-chain Monte Carlo analysis \citep{Lewis:2002ah} to reveal the distinction between different observations. For the current data, we use the ``Planck distance priors" derived from the Planck 2018 data release \citep{Chen:2018dbv}, and the BAO measurements from 6dFGS at $z_{\rm eff} = 0.106$ \citep{Beutler:2011hx}, SDSS-MGS at $z_{\rm eff} = 0.15$ \citep{Ross:2014qpa}, and BOSS-DR12 at $z_{\rm eff} = 0.38$, 0.51, and 0.61 \citep{Alam:2016hwk}. In this work, the fiducial values of cosmological parameters are taken to be the best-fit values of CMB+BAO+SN from \citet{Zhang:2019loq}.
\section{Results and discussion} \label{sec:Result}

\subsection{CMB+FRB}\label{CF}

\begin{table*}[!htb]
\caption{The constraint results of the cosmological parameters in the $w$CDM and CPL models.}
\label{tab:full}
\renewcommand{\arraystretch}{1.5}
\begin{tabular}{cccccccc}
\hline
          Model       & Parameter   & CMB & CMB+BAO& FRB1 & CMB+FRB1& FRB2& CMB+FRB2 \\ \hline
\multirow{4}{*}{$w$CDM} & $\Omega_m$ &
$0.324^{+0.055}_{-0.074}$ & $0.316\pm0.013$ & $0.285^{+0.072}_{-0.047}$ & $0.310^{+0.024}_{-0.030}$& $0.307^{+0.018}_{-0.014}$& $0.312\pm0.012$\\
& $h$ &
$0.676^{+0.061}_{-0.077}$ & $0.674^{+0.013}_{-0.015}$ & $>0.613$ & $0.683\pm 0.030$&$>0.614$&$0.680\pm0.012$\\
& $w$ &
$-1.00^{+0.25}_{-0.22}$ & $-0.995^{+0.061}_{-0.054}$ & $-1.19^{+0.73}_{-0.33}$ & $-1.03\pm 0.10$&$-1.04^{+0.21}_{-0.17}$&$-1.021\pm0.044$\\
& $10^2\Omega_b h^2$ &
$2.235\pm0.015$ & $2.238\pm0.015$ & $2.23\pm0.49$ & $2.235\pm0.013$&$2.25^{+0.64}_{-0.34}$&$2.2353\pm0.0087$\\ \hline
\multirow{5}{*}{CPL} & $\Omega_m$ &
$0.318\pm0.059$ & $0.342^{+0.026}_{-0.029}$ & $0.333^{+0.100}_{-0.076}$ & $0.316^{+0.040}_{-0.054}$&$0.326^{+0.082}_{-0.056}$&$0.314\pm0.023$\\
& $h$ &
$0.682^{+0.052}_{-0.076}$ & $0.650^{+0.024}_{-0.027}$ & $>0.653$ & $0.680\pm0.050$&$>0.647$&$0.679^{+0.023}_{-0.026}$\\
& $w_0$ &
$-0.60\pm0.52$ & $-0.68^{+0.27}_{-0.31}$ & $-0.78\pm0.64$ & $-0.89^{+0.41}_{-0.53}$&$-0.77^{+0.37}_{-0.50}$&$-0.98\pm0.24$\\
& $w_a$ &
$<-0.592$ & $-0.91^{+0.91}_{-0.70}$ & $<-0.557$ & $-0.49^{+1.40}_{-0.92}$&$-1.1^{+2.5}_{-1.5}$&$-0.15^{+0.62}_{-0.54}$\\
& $10^2\Omega_b h^2$ &
$2.236\pm0.015$ & $2.235\pm0.015$ & $2.54^{+1.00}_{-0.72}$ & $2.236\pm0.015$&$2.54^{+1.00}_{-0.63}$&$2.237\pm0.012$\\ \hline
\end{tabular}
\tablecomments{The errors at the 68.3\% confidence level. The results of the cosmological parameters are obtained by using the CMB, CMB+BAO, FRB1, CMB+FRB1, FRB2, and CMB+FRB2 data. FRB1 and FRB2 denote the FRB data in the normal expected scenario (i.e., $N_{\rm FRB}=1000$) and the FRB data in the optimistic scenario (i.e., $N_{\rm FRB}=10,000$), respectively.}
\end{table*}

\begin{figure*}[htb]
\setlength{\abovecaptionskip}{-0.2cm}
\begin{center}
\hspace*{.1cm}
\subfigure[]{\includegraphics[width=0.4\linewidth,angle=0]{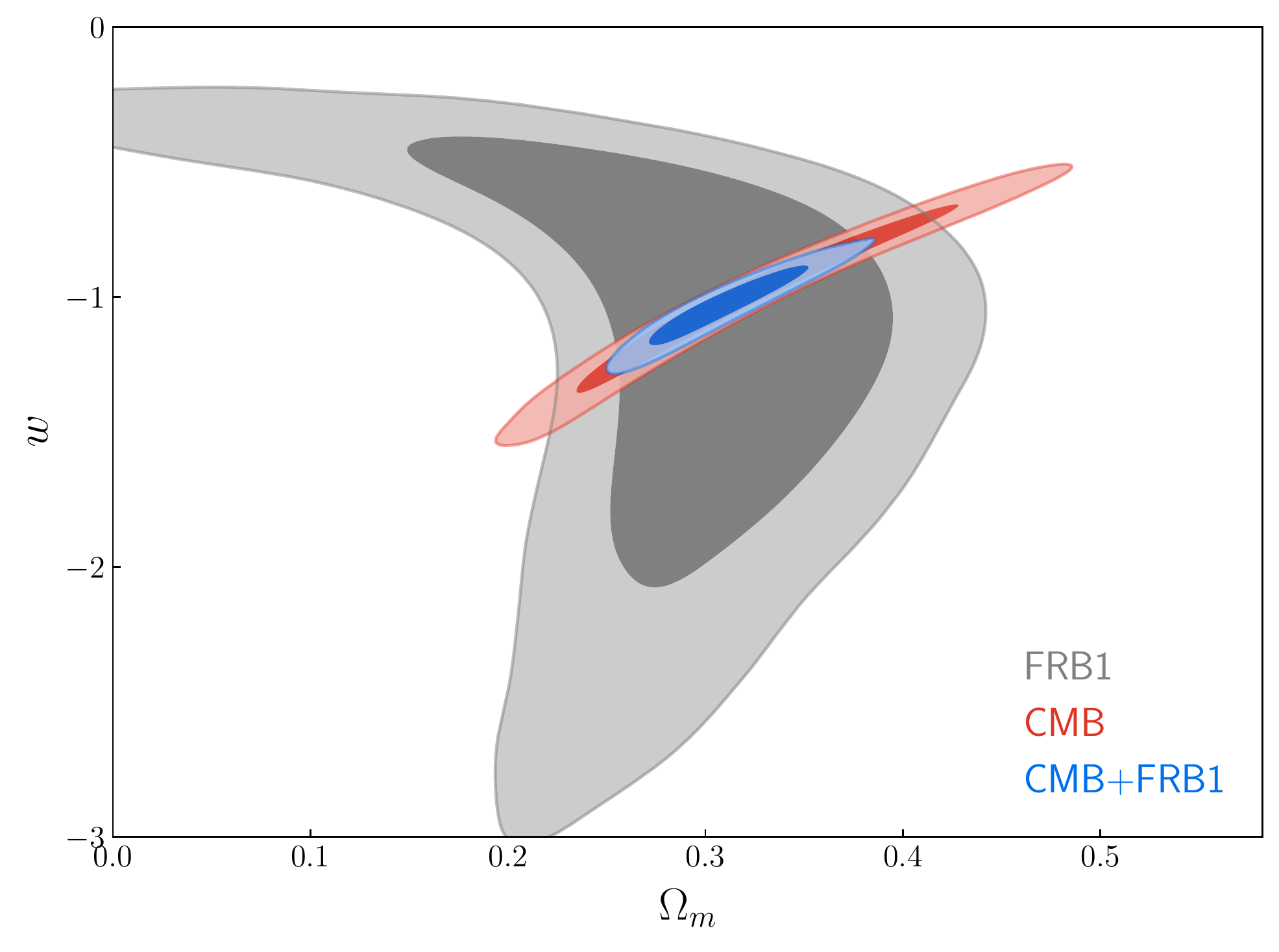}
\label{Fig1.sub.1}}
\hspace*{.1cm}
\subfigure[]{\includegraphics[width=0.4\linewidth,angle=0]{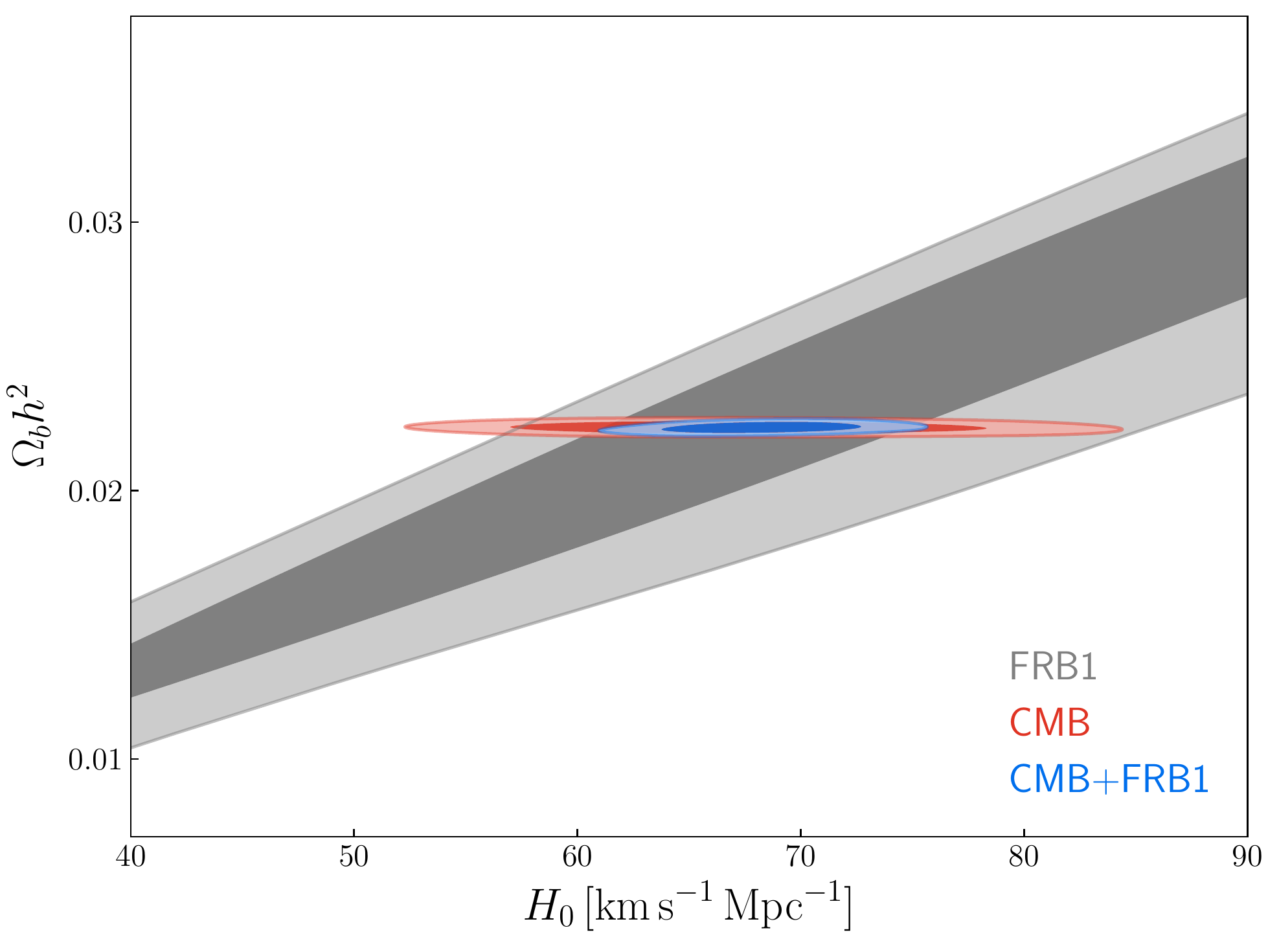}
\label{Fig1.sub.2}}
\end{center}
\caption{Two-dimensional marginalized contours (68.3\% and 95.4\% confidence levels) in the $\Omega_{\rm m}$--$w$ plane (left panel) and the $H_{0}$--$\Omega_{\rm b}h^2$ plane (right panel) for the $w$CDM model, by using FRB, CMB, and CMB+FRB. Here, for the simulated FRB data, the normal expected scenario is assumed.} \label{re}
\end{figure*}

\begin{figure*}[htb]
\setlength{\abovecaptionskip}{-0.2cm}
\begin{center}
\hspace*{.1cm}
\subfigure[]{\includegraphics[width=0.4\linewidth,angle=0]{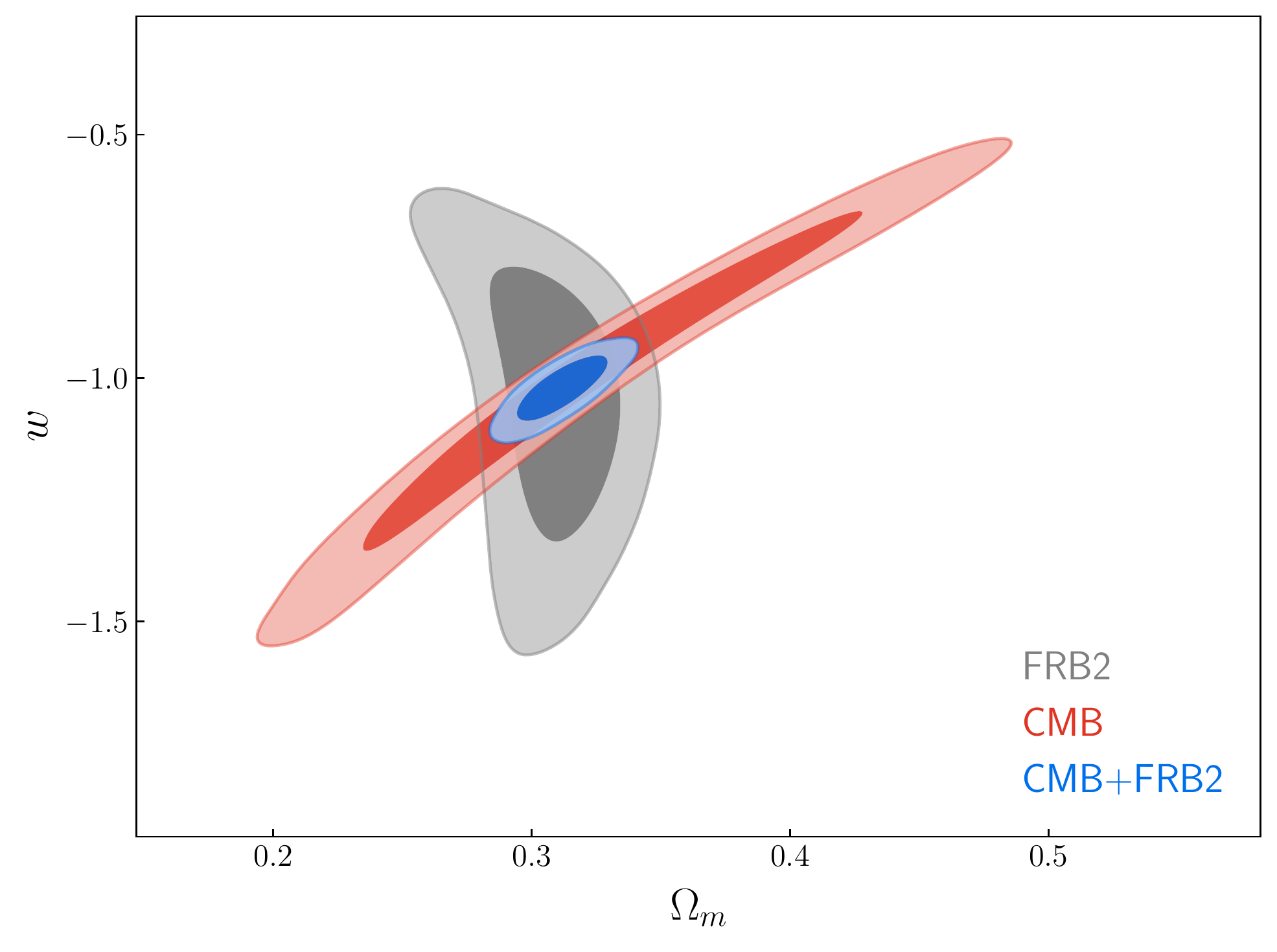}
\label{Fig2.sub.1}}
\hspace*{.1cm}
\subfigure[]{\includegraphics[width=0.4\linewidth,angle=0]{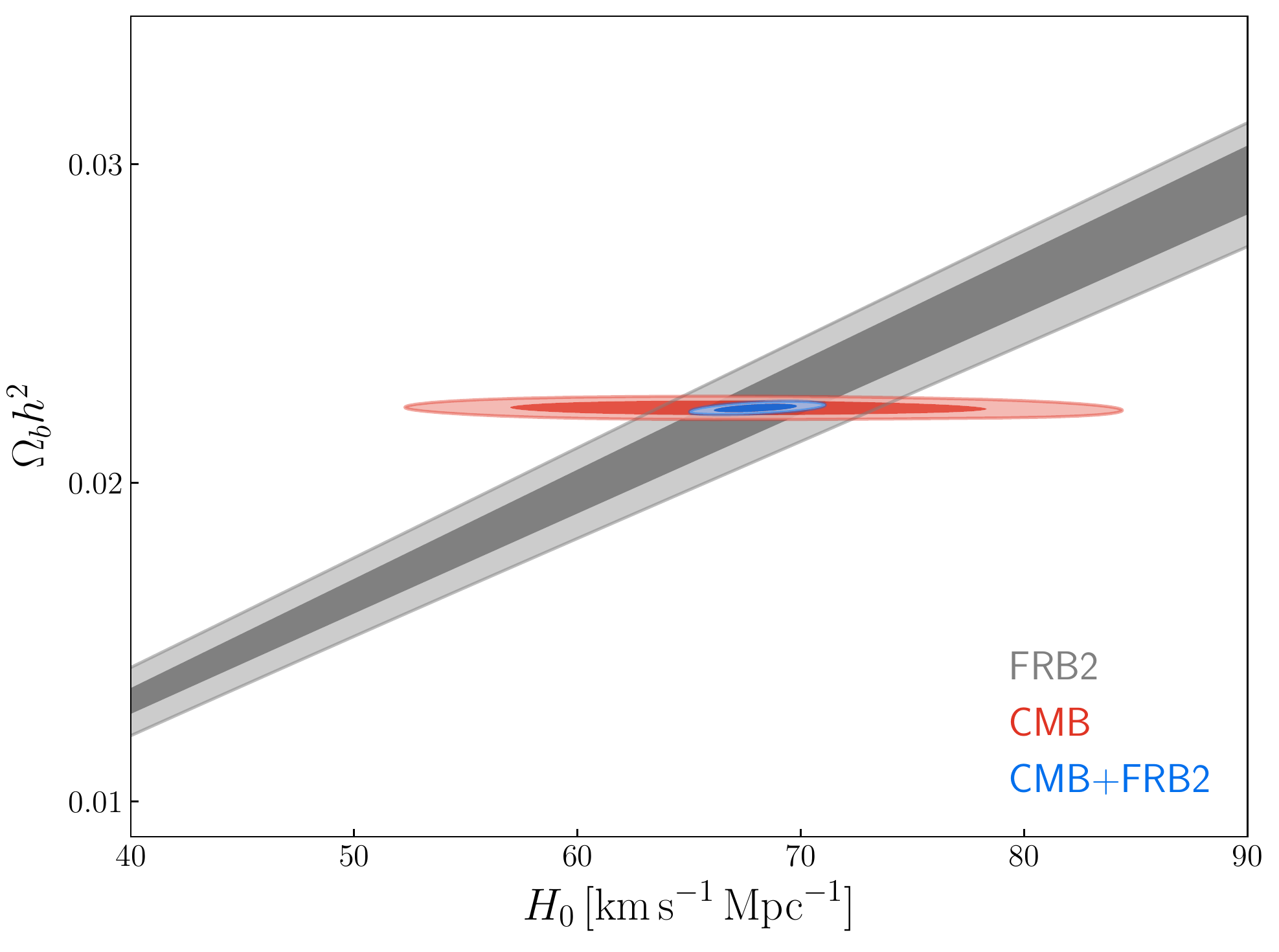}
\label{Fig2.sub.2}}
\end{center}
\caption{Two-dimensional marginalized contours (68.3\% and 95.4\% confidence levels) in the $\Omega_{\rm m}$--$w$ plane (left panel) and the $H_{0}$--$\Omega_{\rm b}h^2$ plane (right panel) for the $w$CDM model, by using FRB, CMB, and CMB+FRB. Here, for the simulated FRB data, the optimistic scenario is assumed.} \label{op}
\end{figure*}

In this subsection, the simulated FRB data are combined with the CMB data to study the help of FRB in cosmological parameter estimation. In Table \ref{tab:full}, we list the best-fit value and the standard 1$\sigma$ error for every cosmological parameter $\xi$ in the $w$CDM and CPL models. In the following, FRB1 and FRB2 denote the FRB data in the normal expected scenario (i.e., $N_{\rm FRB}=1000$) and the FRB data in the optimistic scenario (i.e., $N_{\rm FRB}=10,000$), respectively.

The constraints from the CMB data are tighter than those from the FRB data in the normal expected scenario but weaker than those from the FRB data in the optimistic scenario. We find that the constraints on cosmological parameters are evidently improved for both CMB+FRB1 and CMB+FRB2 combinations. To obtain some insights into how this can be achieved, we plot the two-dimensional marginalized posterior probability distribution contours in the $\Omega_{\rm m}$--$w$ plane for the $w$CDM model in Figure \ref{Fig1.sub.1}, by using FRB1, CMB, and CMB+FRB1. The orientations of the parameter degeneracies formed by CMB and by FRB are rather different, thus the parameter degeneracies are broken by combining the CMB and FRB data. This effect is clearer in Figure \ref{Fig2.sub.1} for the FRB data in the optimistic scenario, in which the FRB data provide a tighter constraint on $\Omega_{\rm m}$ compared to CMB. Quantitatively, the current CMB data combined with the simulated FRB1 and FRB2 data can give the relative errors $\varepsilon(w)=9.7\%$ and $\varepsilon(w)=4.3\%$, respectively, indicating that the constraints are improved by about 59\% and 82\% compared with those using the CMB data alone, respectively.

In Figure~\ref{Fig1.sub.2}, we show the marginalized posterior probability distribution contours in the $H_0$--$\Omega_{\rm b}h^2$ plane for the $w$CDM model. It is obvious that $H_0$ and $\Omega_{\rm b}h^2$ cannot be effectively constrained by FRB alone, since $\mathrm{DM}_{\mathrm{IGM}}$ is proportional to $H_0 \Omega_{\rm b}$ (see Equation (\ref{eq3})). Considering that CMB can constrain $\Omega_{\rm b}h^2$ at high precision, the combination of the CMB and FRB data can break the degeneracy between $H_0$ and $\Omega_{\rm b}h^2$, resulting in a precise measurement on $H_0$. Concretely, the simulated FRB1 data combined with the current CMB data can achieve the relative error $\varepsilon(h)=4.4\%$, indicating a 56\% improvement compared to the one $\varepsilon(h)=10\%$ by using the CMB data alone. The effect of the event number of FRBs is distinct for the data combination CMB+FRB, as can be seen from Figure \ref{Fig2.sub.2}, in which the degeneracy between $H_0$ and $\Omega_{\rm b}h^2$ is extremely strong for the FRB data in the optimistic scenario. Increasing the event number of FRBs from 1000 to 10,000 in the data combination CMB+FRB improves the constraint error on $h$ to 1.8\%, which corresponds to a 82\% reduction in the size of the 1$\sigma$ error of the CMB data.

\begin{figure}[htb]
\includegraphics[width=0.9\linewidth,angle=0]{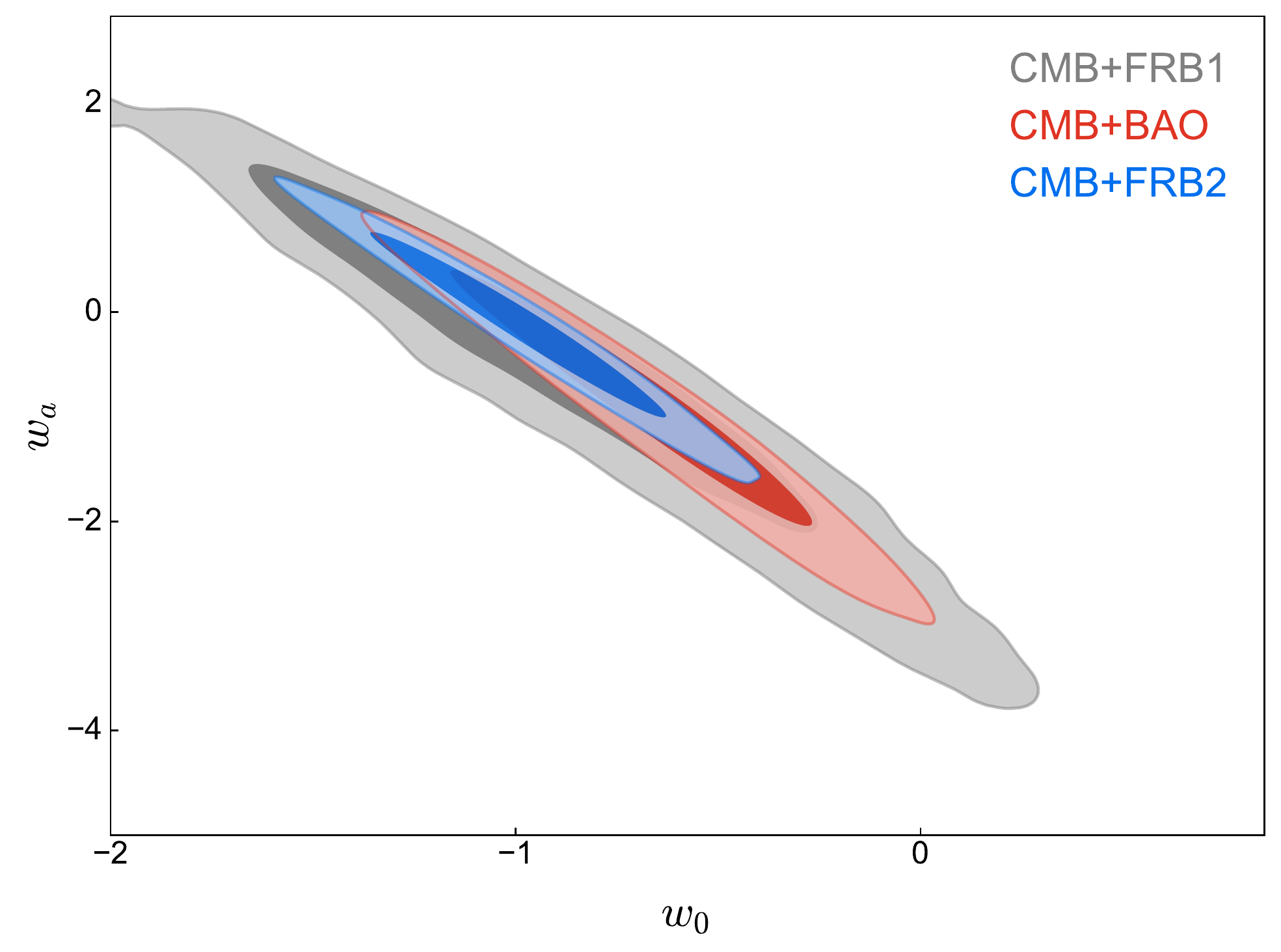}
\caption{Two-dimensional marginalized contours (68.3\% and 95.4\% confidence levels) in the $w_0$--$w_a$ plane for the CPL model, by using CMB+FRB1, CMB+BAO, and CMB+FRB2.} \label{BAO}
\end{figure}



In the last part of this subsection, we compare the capabilities of the BAO and FRB data of breaking the parameter degeneracies inherent in the CMB data (see the fourth, sixth, and eighth columns of Table~\ref{tab:full}). From Figure \ref{BAO} we find that the constraints from CMB+BAO are tighter than those from CMB+FRB1 but weaker than those from CMB+FRB2 in both the $w$CDM and CPL models. For example, the 1$\sigma$ errors on $w_0$ and $w_a$ in the CPL model are 0.47 and 1.16, respectively, by CMB+FRB1, 0.29 and 0.81, respectively, by CMB+BAO, and 0.24 and 0.58, respectively, by CMB+FRB2. In addition, the BAO data can provide little help for the $\Omega_{\rm b}h^2$ constraint compared with CMB alone. However, compared with the result from CMB, the inclusion of the FRB2 data improves the constraint on $\Omega_{\rm b}h^2$ by 42\% in the $w$CDM model.

\subsection{GW+FRB}

\begin{table*}[!htb]
\caption{The 1$\sigma$ errors on the cosmological parameters in the $w$CDM model.}
\label{tab:GW}
\hspace{1.5cm}
\begin{tabular}{cccccc}
\hline
          Model       & Error   & GW & GW+FRB  &  CMB+GW& CMB+GW+FRB \\ \hline
\multirow{8}{*}{$w$CDM} & \multirow{2}{*}{$\sigma(\Omega_m)$} &
\multirow{2}{*}{0.028} & 0.024  & \multirow{2}{*}{0.0067}& 0.0063\\ 
& & & 0.013 & & 0.0057 \\ \cline{2-6}
& \multirow{2}{*}{$\sigma(h)$} &
\multirow{2}{*}{0.013} & 0.012  & \multirow{2}{*}{0.0075}& 0.0070\\ 
& & & 0.0097 & & 0.0062 \\ \cline{2-6}
& \multirow{2}{*}{$\sigma(w)$} &
\multirow{2}{*}{0.18} & 0.16  & \multirow{2}{*}{0.037} & 0.034 \\ 
& & & 0.11 & & 0.030 \\ \cline{2-6}
& \multirow{2}{*}{$10^2\sigma(\Omega_b h^2)$} &
\multirow{2}{*}{...} & 0.022 & \multirow{2}{*}{0.014} &0.012\\
& & &0.013 & & 0.0072 \\ \hline
\end{tabular}
\tablecomments{The errors on the cosmological parameters are obtained by using the GW, GW+FRB, CMB+GW, and CMB+GW+FRB data. The two values in a cell in the columns of GW+FRB and CMB+GW+FRB represent the constraints based on the FRB data in the normal expected scenario and in the optimistic scenario, respectively, from top to bottom.}
\end{table*}

\begin{figure}[htb]
\includegraphics[width=0.9\linewidth,angle=0]{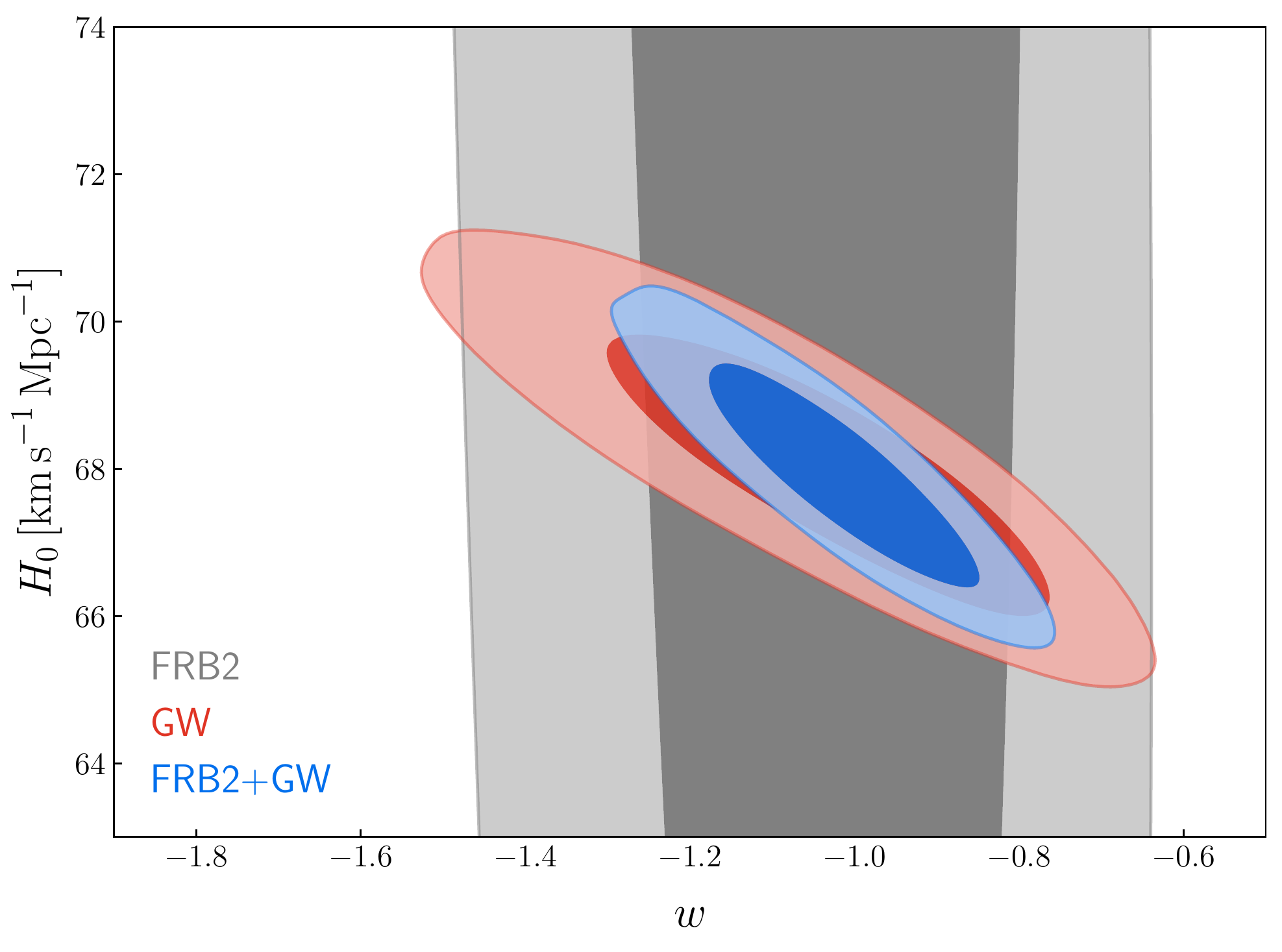}
\caption{Two-dimensional marginalized contours (68.3\% and 95.4\% confidence levels) in the $w$--$H_{0}$ plane for the $w$CDM model, by using FRB, GW, and FRB+GW. Here, for the simulated FRB data, the optimistic scenario is assumed.} \label{GW+FRB}
\end{figure}

In this subsection, we study whether the combination of the simulated FRB and GW data is able to efficiently break the parameter degeneracies. The constraint results in the $w$CDM model are given in Table \ref{tab:GW}.

From Figure \ref{GW+FRB}, only in the optimistic scenario does the data combination GW+FRB evidently improve the constraints on the parameters compared with those using the GW data alone. For example, compared with the results of the GW data alone, the inclusion of FRB2 data reduces the relative error on $\Omega_{\rm m}$ from $9.0\%$ to $4.2\%$. We also find that the data combination GW+FRB2 provides constraints comparable to the CMB+BAO constraints for the parameters $\Omega_{\rm m}$ and $\Omega_{\rm b}h^2$. For the parameters $H_0$ and $w$, we have the constraint errors: $\varepsilon(H_0)=1.4\%$ and $\varepsilon(w)=11\%$ from GW+FRB2, and $\varepsilon(H_0)=2.1\%$ and $\varepsilon(w)=5.8\%$ from CMB+BAO. Compared with the case of CMB+BAO, the combination of the FRB and GW data provides tighter constraint on $H_0$ and looser constraint on $w$. Therefore, the data combination GW+FRB can serve as a low-redshift measure of cosmological parameters, which is fully independent of the CMB and BAO observations.



Although the orientations of the parameter degeneracies formed by FRB and by GW are obviously different, the FRB data cannot provide much help to the GW data, because $H_0$ is poorly constrained using FRB alone. In addition, the standard siren data do not contain information on $\Omega_{\rm b}$, thus they cannot break the strong degeneracy between $H_0$ and $\Omega_{\rm b}h^2$ in FRB. Since we have shown that the combination of CMB and FRB can break the parameter degeneracies in each other, we will further include the CMB data in the data combination in the next subsection.

\subsection{CMB+GW+FRB}

\begin{figure}[htb]
\includegraphics[width=0.9\linewidth,angle=0]{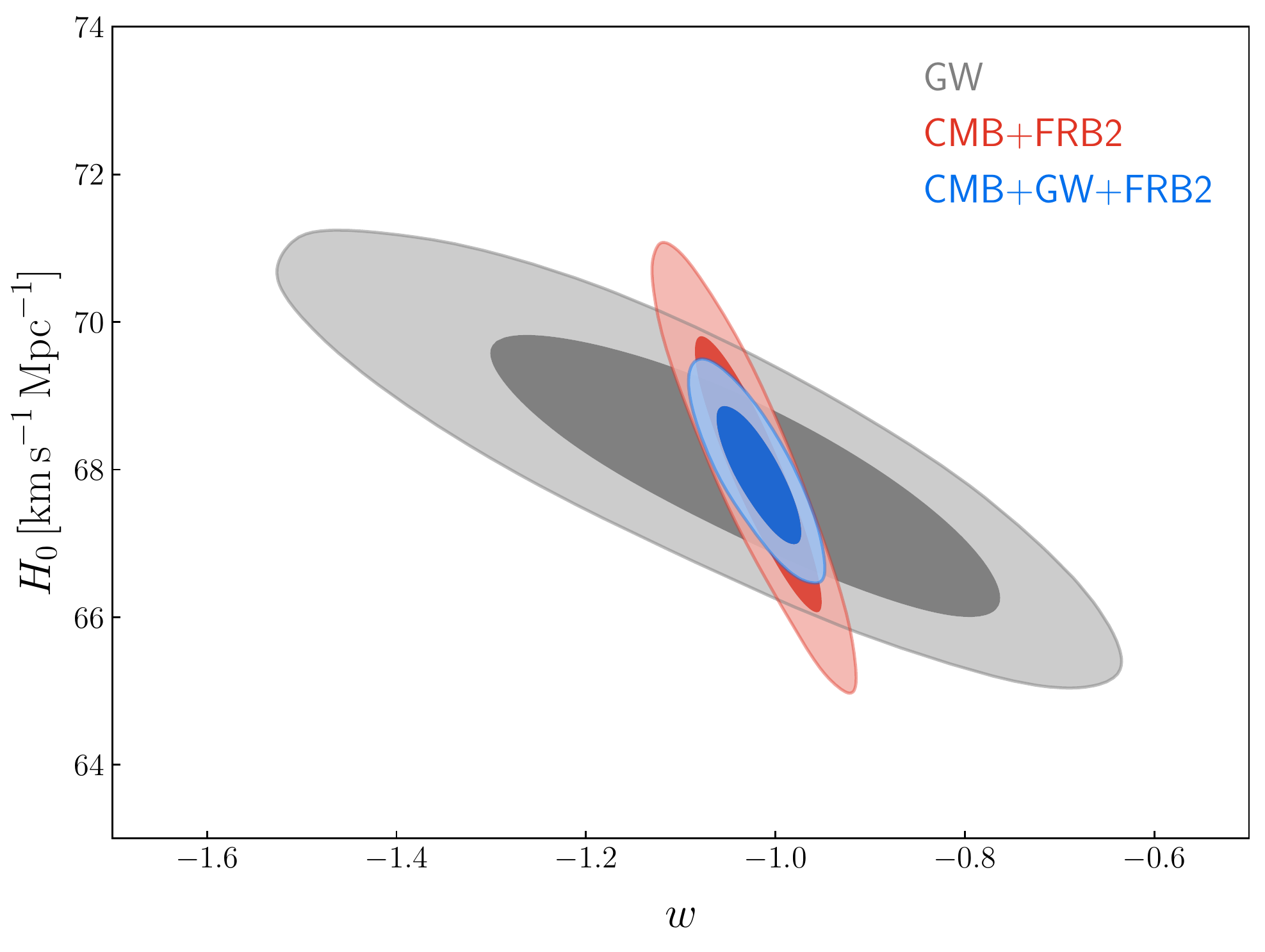}
\caption{Two-dimensional marginalized contours (68.3\% and 95.4\% confidence levels) in the $w$--$H_{0}$ plane for the $w$CDM model, by using GW, CMB+FRB, and CMB+GW+FRB. Here, for the simulated FRB data, the optimistic scenario is assumed.} \label{CMB+GW+FRB1}
\end{figure}

\begin{figure}[htb]
\includegraphics[width=0.9\linewidth,angle=0]{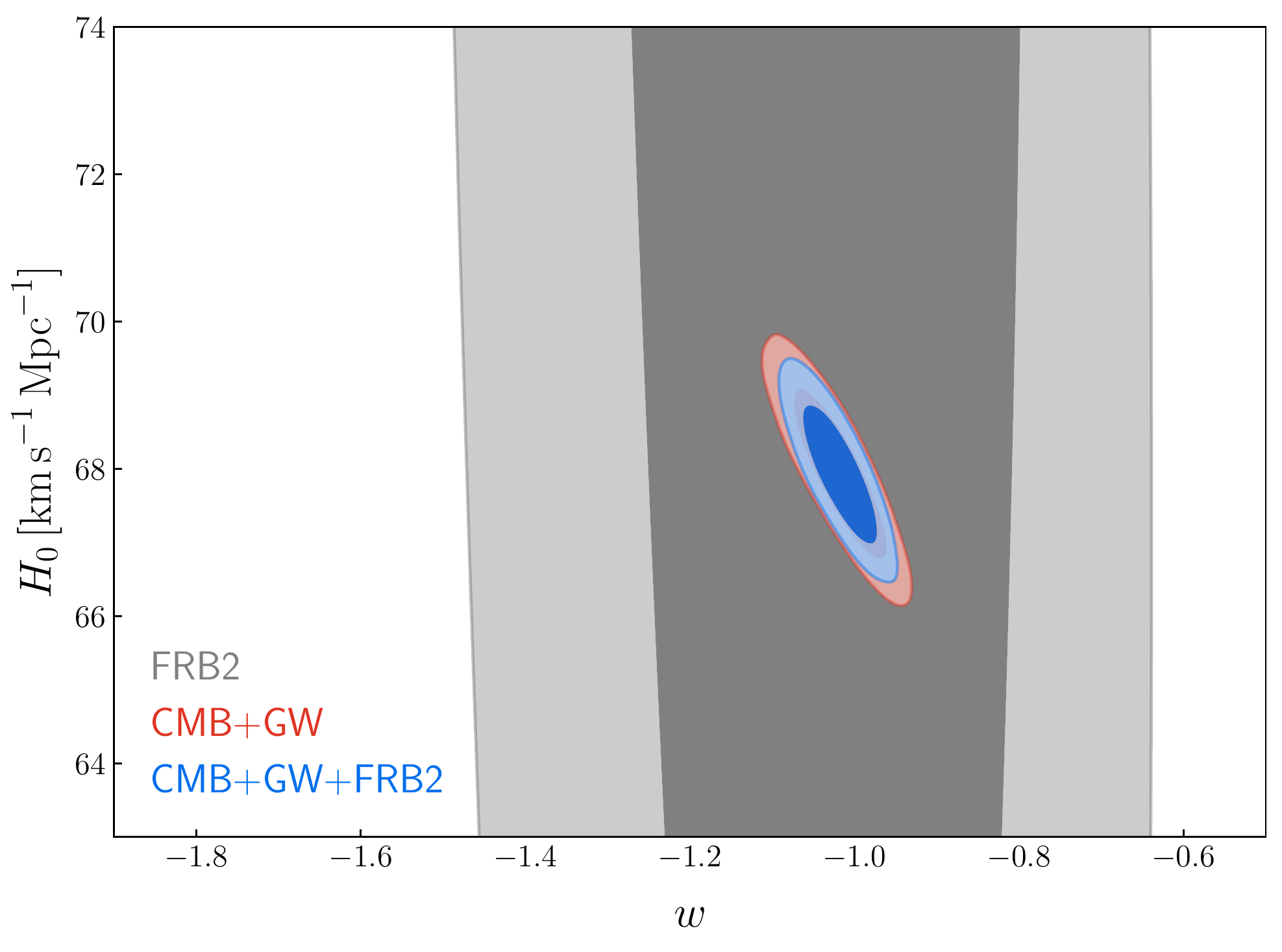}
\caption{Two-dimensional marginalized contours (68.3\% and 95.4\% confidence levels) in the $w$--$H_{0}$ plane for the $w$CDM model, by using FRB, CMB+GW, and CMB+GW+FRB. Here, for the simulated FRB data, the optimistic scenario is assumed.} \label{CMB+GW+FRB2}
\end{figure}

We further investigate the capability of the data combination CMB+GW+FRB to constrain cosmological parameters. We show the constraint contours by using the GW, CMB+FRB2, and CMB+GW+FRB2 data combinations for the $w$CDM model in Figure \ref{CMB+GW+FRB1}, which seems to show that the constraints from the combination CMB+GW+FRB are obviously improved relative to GW and CMB+FRB, due to the different orientations of the degeneracies.

However, from Figure \ref{CMB+GW+FRB2}, it is clear that the constraint results from the data combination CMB+GW+FRB are only slightly better than those from CMB+GW. The constraints from the FRB data are too weak compared to those from CMB+GW. This is also indicated by Figure \ref{op}, in which the parameter degeneracies of the data combination CMB+FRB are determined by CMB, but not by FRB. The precise measurement on $\Omega_{\rm b}h^2$ by CMB leads to a $\mathrm{DM}_{\mathrm{IGM}}$ roughly proportional to $1/H_0$, which is similar to the expression of luminosity distance. Therefore, the FRB data cannot effectively break the parameter degeneracies formed by the GW data.

Although the combination CMB+GW already provides tight constraints on cosmological parameters, the FRB data can still supplement to them. Since the GW data lack information on $\Omega_{\rm b}$, the most improved constraint by including the FRB2 data is given by $\varepsilon(\Omega_{\rm b}h^2)=0.32\%$, which is improved by 49\% compared to the result of CMB+GW. 


\subsection{Discussion}

{Since the progenitors of FRBs have not yet been determined, the redshift distribution and the ${\rm DM_{host}}$ uncertainty are still open issues. In this subsection, we further discuss the dependence of our results on these two factors, and we only show the results in the $w$CDM model using the FRB data in the normal expected scenario as an example.}

\begin{table*}[!htb]
\caption{{The constraint results of the cosmological parameters in the $w$CDM model.}}
\label{tab:dis}
\renewcommand{\arraystretch}{1.5}
\begin{center}{\centerline{
\begin{tabular}{cccccc}
\hline
 Parameter   & FRB3 & CMB+FRB3 & FRB4 & CMB+FRB4 \\ \hline
 $\Omega_m$ &
$0.288^{+0.075}_{-0.051}$ & $0.309\pm0.028$ & $0.281^{+0.078}_{-0.055}$ & $0.308\pm0.031$\\
 $h$ &
$>0.617$ & $0.685^{+0.025}_{-0.035}$ & $>0.590$ & $0.686^{+0.031}_{-0.038}$\\
 $w$ &
$-1.22^{+0.76}_{-0.33}$ & $-1.03^{+0.11}_{-0.09}$ & $-1.29^{+0.87}_{-0.40}$ & $-1.04^{+0.13}_{-0.10}$\\
 $10^2\Omega_b h^2$ &
$2.24\pm0.49$ & $2.235\pm0.014$ & $2.17^{+0.52}_{-0.58}$ & $2.235\pm0.014$\\ \hline
\end{tabular}}}
\end{center}
\tablecomments{{The errors at the 68.3\% confidence level. The results of the cosmological parameters are obtained by using FRB3, CMB+FRB3, FRB4, and CMB+FRB4. Here, \emph{FRB3} and \emph{FRB4} denote the FRB data with the redshift distribution tracking SFH and the FRB data with the ${\rm DM_{host}}$ uncertainty $\sigma_{\rm host} = 150\, {\rm pc\,cm^{-3}}$, respectively.}}
\end{table*}

{First, we further consider another scenario in which the FRB redshift distribution follows the star-formation history (SFH) \citep{Caleb}. In this case, the SFH-based redshift distribution function of FRBs is \citep{Munoz:2016tmg},
\begin{align}
N_{\rm SFH}(z)={\cal N_{\rm SFH}} \dfrac{\dot \rho_* (z) \,{d^2_{\rm C}}(z)}{H(z)(1+z)} e^{-d_L^2(z)/[2 d_L^2(z_{\rm cut})]},
\end{align}
where the parameterized density $\dot \rho_* (z)$ reads
\begin{align}
\dot \rho_* (z)= l \dfrac{a+b z}{1+\left(z/c\right)^d},
\end{align}
with $a=0.0170$, $b=0.13$, $c=3.3$, $d=5.3$, and $l=0.7$ \citep{Cole,Hopkins}, and $\cal N_{\rm SFH}$ is a normalization factor.

The results are shown in Table \ref{tab:dis}, where we use FRB3 to represent the FRB data whose distribution tracks SFH. We find that, for both the case using the FRB data alone and the case using the data combination FRB+CMB, the constraints based on the FRB data tracking the SFH redshift distribution are not obviously different from those based on the FRB data assuming a constant comoving density. For example, for the DE EoS parameter $w$, we have the constraint errors: 44.5\% from FRB1, 44.7\% from FRB3, 9.7\% from CMB+FRB1, and 9.5\% from CMB+FRB3. Therefore, the FRB redshift distribution almost has no impact on the cosmological parameter estimation in our work.

Second, the ${\rm DM_{host}}$ has limited theoretical motivation and is difficult to estimate accurately for every event. Therefore, we further set the ${\rm DM_{host}}$ uncertainty $\sigma_{\rm host} = 150\, {\rm pc\,cm^{-3}}$ as a conservative scenario and discuss its impact on the cosmological parameter estimation. The results are shown in the fourth and fifth columns of Table \ref{tab:dis}, where FRB4 denotes the FRB data with $\sigma_{\rm host} = 150\, {\rm pc\,cm^{-3}}$.

We find that the constraints in the conservative scenario (i.e. $\sigma_{\rm host} = 150\, {\rm pc\,cm^{-3}}$) are slightly looser than those in the normal expected scenario (i.e. $\sigma_{\rm host} = 30\, {\rm pc\,cm^{-3}}$). To be specific, the simulated FRB4 data and the data combination CMB+FRB4 can give the relative errors $\varepsilon(w)=49.2\%$ and $\varepsilon(w)=11.1\%$, respectively, which are 10.6\% and 14.4\% larger than those from the FRB1 data and the data combination CMB+FRB1, respectively. From Eq. (\ref{eq6}), $\sigma_{\rm host}$ contributes to the total uncertainty of $\rm{DM}_{\rm{IGM}}$ with a factor $1/(1+z)$, so it can only slightly impact FRB's capability to constrain the cosmological parameters and break the degeneracies between the parameters, despite the $\sigma_{\rm host}$ changing a lot. Through the analysis above, we believe that some other factors may affect the quantitative constraints, but the main conclusions will still hold.}

\section{Conclusion} \label{sec:con}
In this work, we study the capability of future FRB data to improve the cosmological parameter estimation. For the FRB
event numbers, we consider a normal expected scenario, i.e., $N_{\rm FRB}=1000$, and an optimistic scenario, i.e., $N_{\rm FRB}=10,000$, as examples. We also consider two dynamical dark energy cosmological models, i.e., the $w$CDM and CPL models.

We find that although the FRB data alone cannot effectively constrain $H_0$ and $\Omega_{\rm b}h^2$, the combination of the current CMB data and the simulated FRB data can provide rather good constraints on the Hubble constant $H_0$ and dark energy parameters, due to the fact that FRB is able to break the parameter degeneracies inherent in the CMB data. Both $H_0$ and the EoS of dark energy $w$ in the $w$CDM model are improved by about 50\% by including the FRB data in the normal expected scenario, and are improved by about 80\% by including the FRB data in the optimistic scenario, compared with those using the CMB data alone. For both the $w$CDM and CPL models, the optimistic FRB data have a better capability, compared with the BAO data, of breaking the parameter degeneracies inherent in the CMB data. Compared with the BAO data, the advantage of the FRB data resides in the better constraint on $\Omega_{\rm b}h^2$.

We also investigate the capability of the combination of the FRB data and the GW standard siren data, another future low-redshift cosmological probe, to estimate cosmological parameters. We find that the constraints on cosmological parameters from the data combination GW+FRB are comparable to those from CMB+BAO. Thus, this combination may provide a novel low-redshift probe of the cosmological parameters, independent of the CMB and BAO data. If we further include the current CMB data in the data combination, the contributions to the constraints of the combination CMB+GW+FRB mainly come from CMB+GW. In fact, the bottleneck of the FRB data in cosmology is the strong degeneracy between $H_0$ and $\Omega_{\rm b}h^2$. In this case, even though the data combination CMB+GW provides tight constraints, the inclusion of FRB data can still improve the constraint on $\Omega_{\rm b}h^2$ evidently.

{Finally, we evaluate the effects of the FRB redshift distribution and the ${\rm DM_{host}}$ uncertainty on the constraint results. The FRB redshift distribution following SFH has little impact on the cosmological parameter estimation. A larger ${\rm DM_{host}}$ uncertainty certainly makes the constraint results worse, but the major conclusions still hold even in a rather conservative scenario. }For now, the FRB cosmology is still an open topic, but we believe that future plentiful FRB observations will play a significant role in cosmological parameter estimation, to supplement and verify the results from other observations.

\acknowledgments
We are very grateful to Ling-Feng Wang for fruitful discussions. This work was supported by the National Natural Science Foundation of China (Grants Nos. 11975072, 11875102, 11835009, 11690021, 11722324, 11690024, and 11920101003), the Liaoning Revitalization Talents Program (XLYC1905011), the Fundamental Research Funds for the Central Universities (N2005030 and N180503014), and the National Program for Support of Top-Notch Young Professionals (W02070050).

\end{document}